\pdfoutput=1
\documentclass[11pt]{article}
\usepackage[pdftex, hidelinks]{hyperref}
\usepackage{acl}
\usepackage{amssymb}
\usepackage{times}
\usepackage{latexsym}
\usepackage{amsmath}
\usepackage[T1]{fontenc}
\usepackage[utf8]{inputenc}
\usepackage{microtype}
\usepackage{inconsolata}
\usepackage{graphicx}
\usepackage{multirow}
\usepackage{booktabs}
\usepackage{breqn}
\title{Enhancing Code LLMs with Reinforcement Learning in Code Generation: A Survey}
\author{
Junqiao Wang$^{*1}$, Zhangzeng$^{*3}$, Yangfan He$^{*2}$, Zihao Zhang$^{1}$,  Xinyuan Song$^{13}$\\ Yuyang Song$^{1}$, Tianyu Shi$^{4\dagger}$, Yuchen Li$^{6}$, 
Hengyuan Xu$^{1}$, Kunyu Wu$^{1}$, Xin Yi$^{8}$, \\ Zhongwei Wan$^{7}$, Xinhang Yuan$^{1}$, Zijun Wang$^{15}$, Kuan Lu$^{9}$, Menghao Huo$^{11}$,\\
Jingqun Tang$^{14}$, Guangwu Qian$^{1}$, Keqin Li$^{12}$, Qiuwu Chen$^{6}$, Lewei He$^{3}$ \\
$^{1}$SCU,  
$^{2}$UMN,  
$^{3}$SCNU,  
$^{4}$UofT,  
$^{5}$Henan Runtai Tech,
$^{6}$AIGCode,  
$^{7}$OSU,  
$^{8}$NJU, \\
$^{9}$Cornell,  
$^{10}$WashU, 
$^{11}$SCU,  
$^{12}$AMA University, 
$^{13}$Emory,  
$^{14}$ByteDance,
$^{15}$UCSC \\ \texttt{tianyu.s@outlook.com}  
$^{*}$Equal contribution $^{\dagger}$Corresponding author} 
\usepackage{graphicx}

\begin{document}
\maketitle
\vspace{300cm}
\begin{abstract}
With the rapid evolution of large language models (LLM), reinforcement learning (RL) has emerged as a pivotal technique for code generation and optimization in various domains. This paper presents a systematic survey of the application of RL in code optimization and generation, highlighting its role in enhancing compiler optimization, resource allocation, and the development of frameworks and tools. Subsequent sections first delve into the intricate processes of compiler optimization, where RL algorithms are leveraged to improve efficiency and resource utilization. The discussion then progresses to the function of RL in resource allocation, emphasizing register allocation and system optimization. We also explore the burgeoning role of frameworks and tools in code generation, examining how RL can be integrated to bolster their capabilities. This survey aims to serve as a comprehensive resource for researchers and practitioners interested in harnessing the power of RL to advance code generation and optimization techniques.
\end{abstract}

\section{Introduction}
\begin{figure}
    \centering
    \includegraphics[width=\linewidth]{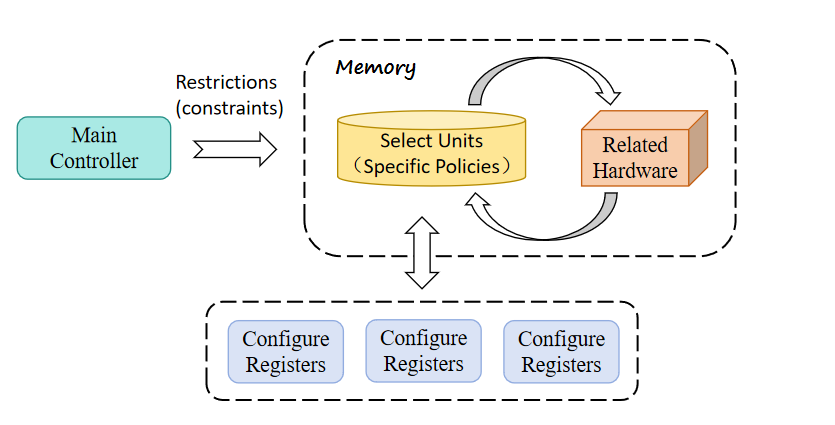}
    \caption{The controller is in control of the resources.}
    \label{fig:fig11}
\end{figure}

As software systems grow more complex with tighter development timelines, manual code development and optimization become impractical to a certain extent. Code generation and optimization from natural language (NL) have thus become essential for boosting software development efficiency \cite{nl2codesurvey, nl2codesurvey2}. Meanwhile, advances in natural language processing (NLP), particularly in large language models (LLMs), have opened new possibilities for code generation. Compiler optimizations are essential in enhancing software performance and reducing resource consumption. Conventional compiler optimization relies on techniques like autotuning \cite{autotuning1}, while deep learning approaches for optimized compiler sequences \cite{Li2020TheDL} struggle with generalization. Although large language models (LLMs) improve code generation and optimization \cite{Cummins2023LargeLM}, they often produce biased or inconsistent outputs \cite{wang-etal-2023-recode,Grounded-Copilot} and require time-consuming pre-training with code-specific models like Code T5 \cite{wang-etal-2021-codet5} and Code T5+ \cite{wang-etal-2023-codet5}. In Figure \ref{fig:fig11}, we present a schematic diagram of memory management, wherein the main controller selects specific strategies based on the constraints. Subsequently, it interacts with relevant hardware components such as the compiler and configures the registers, thereby achieving an overall optimization effect. 

\begin{figure*}
    \centering
    \includegraphics[width=0.62\linewidth]{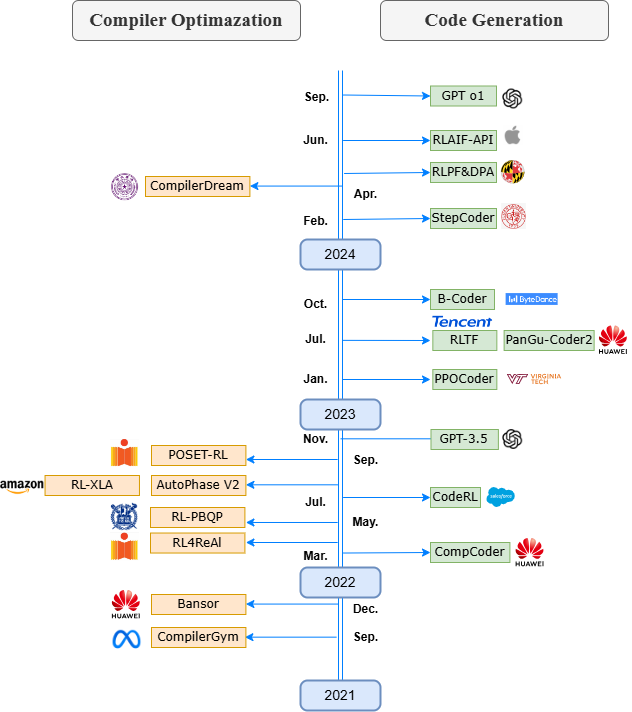}
    \caption{The diagram illustrates the development of code optimization and generation from 2021 to 2024, covering key projects such as CompilerDream, the GPT series, and other important contributions from various organizations.}
    \label{fig:fig22}
\end{figure*}

Current Code LLM research is more focused on the pre-training of code-related corpora. Reinforcement learning (RL), as a method that can learn the optimal strategy in complex environments, provides a new approach to code generation \cite{13}, and optimization \cite{Bendib2024ARL} in Code LLMs. It allows label-free input-output pairs and leverages existing knowledge and refining strategies through trial and error. The advantages of using RL in code optimization and compiler enhancement lie in its capacity to reduce reliance on pre-trained models and to enable large language models (LLMs) to adapt more flexibly to evolving environmental conditions. Figure \ref{fig:fig22} illustrates related work on reinforcement learning applied to compiler optimization and code generation.

Given the great potential for reinforcement learning applications in the most important aspects of improving software performance and efficiency, this paper explores how reinforcement learning can be successfully applied and provides a broader overview of RL-related issues to encourage more researchers to benefit from advances in RL.

\section{Background and Fundamentals}
\label{section:2}

\subsection{Code Generation: Concepts and Evolution}
 Code generation is a fundamental task of Code LLMs, which is essential for automatic programming tasks by generating executable code from natural language descriptions \cite{4}. These descriptions usually contain statements of programming problems and sometimes include information about the programming context, such as function signatures or assertions, and also a formal input and output. The generated code is then executed by a compiler or interpreter and verified by unit tests to ensure that the generated code meets the requirements and works correctly.
 
 Although LLMs have made significant progress in code generation, there are still some challenges in the quality of the code. Studies have shown that LLM can produce a shorter but more complex code when dealing with complex problems, which is a discrepancy compared to standard solutions \cite{Dou2024WhatsWW}.

As the LLM's context learning capabilities advance, sample code can be introduced into the code generation process to enhance the generation or to control the code format. These examples consist of a fixed set of example pairs that contain several pairs of example inputs and corresponding output codes. By including these examples, the model can refer to similar pairs of inputs and outputs during the generation process, thus improving the accuracy and consistency of the generated code.
Decoding strategies commonly used in code generation include two main categories: deterministic strategies and sampling strategies. Deterministic strategies include greedy search and bundle search, which seek to generate the optimal solutions. In contrast, sampling strategies employ methods such as temperature sampling, Top-K sampling, and Top-P (kernel) sampling to introduce variety and flexibility for a wide range of possible code solutions. These different decoding strategies provide a variety of implementations for code generation and adapt to different application requirements.

\subsection{Reinforcement Learning in Code LLMs}

Reinforcement learning (RL) is a technique that determines the best strategies by receiving reward signals from its environment interactions \cite{OPDR:19}. Its goal is to discover an optimal policy parameter $\theta$ that maximizes the sum of rewards through ongoing engagement with the environment \cite{37}. The distribution $\pi_\theta(a|s)$ indicates the probability of choosing action $a$ given state $s$. Direct computation of cumulative rewards is challenging due to the environment being frequently unknown or only partially observable. To address this, value-based and policy-based methods are used to approximate cumulative rewards or gradients, facilitating iterative updates of $\theta$.
Within the realm of reinforcement learning applied to Code LLMs, policy-based methods, such as the PPO method and the Actor-Critic framework, are particularly significant. The Actor-Critic architecture is widely employed in reinforcement learning by combining an action executor (actor) and an evaluator (critic) 
 \cite{Konda1999ActorCriticA}. The actor's responsibility is to execute actions following the present policy, while the critic assesses the actions' values and provides feedback to enhance the actor's strategy. 
 
 PPO enhances the strategic model by training a value function and incorporating token-wise KL penalties into the rewards to balance the updates to the strategy and prevent excessive optimization of the reward model \cite{Ouyang2022TrainingLM}. The value function is frequently distinct and comparable in size to the policy model, which can result in significant computational and memory requirements. Moreover, within reinforcement learning (RL), the value function serves as a baseline for computing advantages and reducing variance. However, in the context of large language models (LLMs), the reward model generally computes the reward only for the final token, which can complicate the training of the value function for individual tokens. 
 
Some research has introduced DPO \cite{rafailov2024directpreferenceoptimizationlanguage} and GRPO \cite{shao2024deepseekmathpushinglimitsmathematical} approaches to address the problems discussed. Direct preference Optimization (DPO) fundamentally focuses on directly optimizing policies to match human preferences, avoiding the requirement of policy enhancement via reinforcement learning (RL) \cite{rafailov2024directpreferenceoptimizationlanguage}. This method eliminates the reliance on reward models in traditional RL, instead opting to fit an implicit reward model using a simple classification objective, from which the optimal policy can be articulated. GRPO takes a different approach by removing the extra value function and directly incorporating the KL divergence between training and reference policies into the loss function \cite{shao2024deepseekmathpushinglimitsmathematical}. It uses the average reward from different solutions to the same problem as a baseline, streamlining the PPO training process, and mitigating the risk of excessively optimizing model rewards.

In summary, reinforcement learning (RL) improves Code LLMs by tuning policy parameters to increase rewards. It employs methods such as PPO, DPO, and GRPO to enhance strategies via environmental interactions and produce code that aligns with human preferences.

\subsection{RL-based Fine-tuning Algorithms in LLMs}
Reinforcement Learning from Human Feedback (RLHF) has become a crucial algorithmic strategy. Using feedback from humans, RLHF fine-tunes large language models, aligning their outputs with human preferences or specific task objectives. This approach is especially significant in complex tasks, such as program synthesis, where traditional supervised learning struggles to grasp subtle performance indicators. In RLHF, models are trained to favor actions that receive the most favorable evaluation from humans, allowing greater precision in controlling language and code generation.

\begin{figure}
    \centering
    \includegraphics[width=\linewidth]{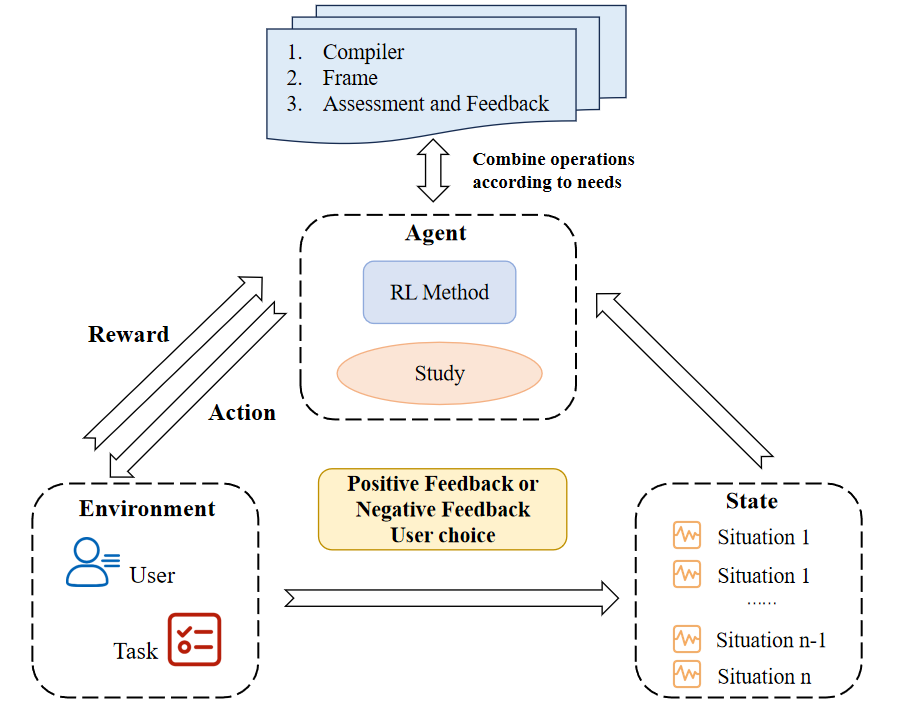}
    \caption{Principles of reinforcement learning (RL) code generation.}
    \label{fig:fig33}
\end{figure}

In the context of code generation, Reinforcement Learning from Human Feedback (RLHF) encounters specific hurdles, as achieving functional correctness—a pivotal aim in programming—cannot be reliably accomplished with token-based similarity metrics such as BLEU or ROUGE, which are typically utilized in translation and summarization tasks. In code generation, token similarity does not consistently correlate with correctness or functionality. Consequently, it is crucial to employ reward signals that directly assess program correctness. Unit test signals offer a potent solution here: they provide a concrete measure of functionality, as programs that pass unit tests can be deemed functionally correct. The feedback system of the reinforcement learning (RL) framework is depicted in Figure \ref{fig:fig33}, through which various elements are utilized to evaluate the action and provide feedback to optimize the agent behavior with diverse operational strategies. Users can also select preferred results to influence the outcomes, enabling RL to exhibit flexibility and adaptability in dynamic environments. By exploiting these unit test outcomes as reward signals, RL-based methods can bring code generation models more aligned with desired outputs, thereby narrowing the gap between generated code and actual functional requirements.

Rather importantly, reinforcement learning (RL)-based fine-tuning methods are crucial in code generation tasks. These methods often employ execution-guided synthesis techniques to refine the strategy of the code model by detailed tuning, which ensures that the generated code is both correct and functionally meets expectations. For instance, this process might involve conducting real-time functional tests on code produced by the model, then adjusting the model's behavior according to the outcomes, thereby enhancing the model's capability to handle intricate programming tasks.

\section{Frameworks in Code Generation and Optimization}
\label{section:4}
\subsection{Theoretical Basis of Code Generation and Optimization}
Generating code entails transforming a natural language description into source code. Given a natural language input detailed as a sequence $x = [x_1, \ldots, x_{|x|}]$, a language model (LM) $p_{LM}$ is used to predict the next token sequentially. At each time step $t$, the LM calculates the probability distribution for the following token, considering all previous tokens, represented as $p_{LM}(x_t \mid x_{1:t-1})$. The probability of creating a program $y$ consisting of the token sequence $y = [x_{|x|+1}, \ldots, x_{|x|+|y|}]$ is computed as the product of these conditional probabilities:
\begin{equation}
P(y \mid x) = \prod_{t=|x|+1}^{|x|+|y|} p_{LM}(x_t \mid x_{1:t})
\end{equation}
Within the framework of few-shot learning utilizing large language models (LLMs), the generation process frequently relies on a predetermined ensemble of $m$ exemplars, represented by $\left\{ \left\langle x_i, y_i \right\rangle \right\}_{i \leq m}$. As a result, the code generation via LLM can be described as:
\begin{equation}
P_{LM}(y \mid x) = P\left(y \mid x, \left\{ \left\langle x_i, y_i \right\rangle \right\}_{i \leq m} \right)
\end{equation}       
Optimizing code involves substituting equivalent code to enhance efficiency in terms of time and space. Local optimization focuses on regions with high time complexity to boost code performance, whereas global optimization considers the overall code structure and its execution. With technological progress, optimization also occurs during the compilation phase, including intermediate code optimization, which refines code structure, and object code optimization, which transforms intermediate code into effective machine code. Additionally, dynamic optimization happens during the program's runtime. Optimizing algorithms and data structures markedly diminishes computational and spatial complexities. The equation to determine the optimal model parameters, \(\theta^*\), is given by: 
\begin{equation}
\theta^* = \arg\max_{\theta} P(y_{\text{best}} \mid x; \theta)
\end{equation}
In this context, \(\theta^*\) stands for the set of parameters that optimizes the likelihood of producing the best candidate program \(y_{\text{best}}\) from the input program \(x\) using the model parameters \(\theta\).

\subsection{Pre-training and Post-training}
\subsubsection{Construction of Datasets}
\begin{figure}
    \centering
    \includegraphics[width=\linewidth]{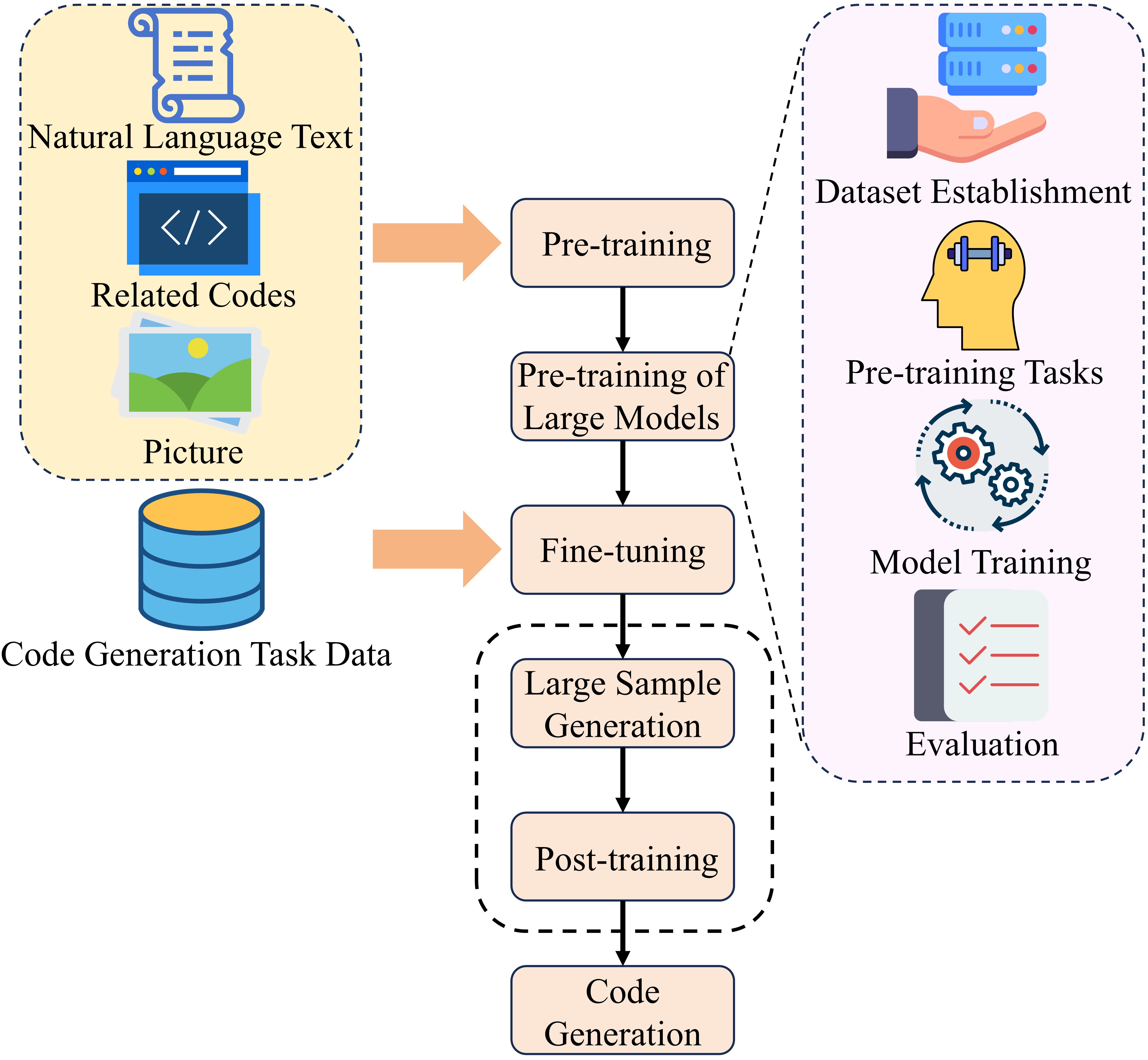}
    \caption{Flowchart of training a code language model (Code LLM).}
    \label{fig:fig44}
\end{figure}

At the outset of constructing datasets for large language models (LLMs) oriented towards code, it is crucial to define the objectives and requirements of the dataset clearly. For such models, the core purpose of the dataset is to enhance the model's capabilities in code generation, comprehension, and task execution. Therefore, collecting code data that encompasses rich algorithmic logic, adheres to good programming practices, and aligns with practical application scenarios becomes particularly important. Figure \ref{fig:fig44} illustrates the training process for code language models (Code LLM), proceeding through data preprocessing, model pre-training, fine-tuning, post-training, and other operations for the applications in code generation. 

To obtain high-quality data sources, GitHub, a widely recognized platform for code sharing and collaboration, serves as an ideal original data source. This platform hosts a vast array of open-source projects, providing us with abundant code data. When selecting data, emphasis should be placed on choosing code files that are logically complete, standardly formatted, and clearly commented. Additionally, to further diversify the dataset, the Common Crawl dataset, a large-scale web scraping dataset, is often utilized. By parsing code-related content from it, a more varied set of code data can be obtained.

However, raw data often contains unrecognized or duplicated content, making data preprocessing an indispensable step. Initially, a rigorous data cleaning process should be implemented to remove non-informative content such as pure hexadecimal code and overly short code snippets, with filtering based on generic attributes like file size and number of lines. Heuristic filtering rules are applied for more refined cleaning tailored to the data characteristics. Given the high degree of repetition in GitHub source code, file-level deduplication strategies, including exact and fuzzy deduplication, are typically employed to effectively reduce data redundancy.

Based on data cleaning, further quality optimization is conducted on the code files. Firstly, code files are checked for their ability to run independently, ensuring they do not depend on external files or libraries. Secondly, code files with unclear logic or disordered structure are eliminated, ensuring the code adheres to standard formatting norms and is converted into a format suitable for LLM training.

To further enhance model performance, during the data refinement stage, not only the cleaned raw data is used, but also algorithm corpora and synthetic data are introduced. These high-quality data rewrites aid the model in better memorizing and embedding knowledge. While preserving the integrity of the original data distribution, downsampling is applied to data from certain high-resource programming languages to improve data utilization efficiency. Simultaneously, to bolster the model's instruction comprehension capabilities, large-scale instruction data is synthesized. Language sampling, task specification modules, and other means are employed to ensure the diversity and specificity of the instruction data. Throughout the construction process, a combination of manual inspection and perplexity (PPL)-based evaluation methods is used to comprehensively assess the data's quality and learnability \cite{huang2024opencoder}.

On the other hand, addressing the limitations of traditional datasets, researchers have proposed the RL4RS framework, a novel system evaluation framework. This framework encompasses multiple dimensions such as environmental simulation evaluation, environmental evaluation, counterfactual policy evaluation, and opportunity test set construction, aiming to provide a more comprehensive assessment of RL algorithm performance. Unlike most related research, RL4RS evaluates policies directly on raw data, avoiding the impact of environment model generalization errors. By introducing counterfactual policy evaluation algorithms and real-world datasets, RL4RS significantly enhances its alignment with the real world \cite{wang2023rl4rs}.

As dataset size continues to expand, the time and labor required to obtain high-quality human feedback become technological bottlenecks. To address this issue, researchers have introduced ULTRAFEEDBACK \cite{cui2023ultrafeedback}, a large-scale, high-quality, and diversified AI feedback dataset. This dataset broadens the scope and depth of instructions and responses, covering a wider range of user-assistant interaction scenarios while mitigating symbolic bias to improve AI feedback reliability. However, large language models face numerous challenges when implementing Reinforcement Learning from Human Feedback (RLHF) \cite{dong2024rlhf}, such as introducing additional PTX losses and studying the distribution shifts between Supervised Fine-Tuning (SFT) and RLHF. To tackle these challenges, SFT data can be used as alternative data for fine-tuning, enhancing transparency from pretraining to SFT and fostering a better understanding of model changes. Furthermore, resource considerations are also crucial in large language model training. High-performance GPUs are used for training, and a combination of human and AI approaches is employed to improve testing consistency.

\begin{figure*}
    \centering
    \includegraphics[width=\textwidth]{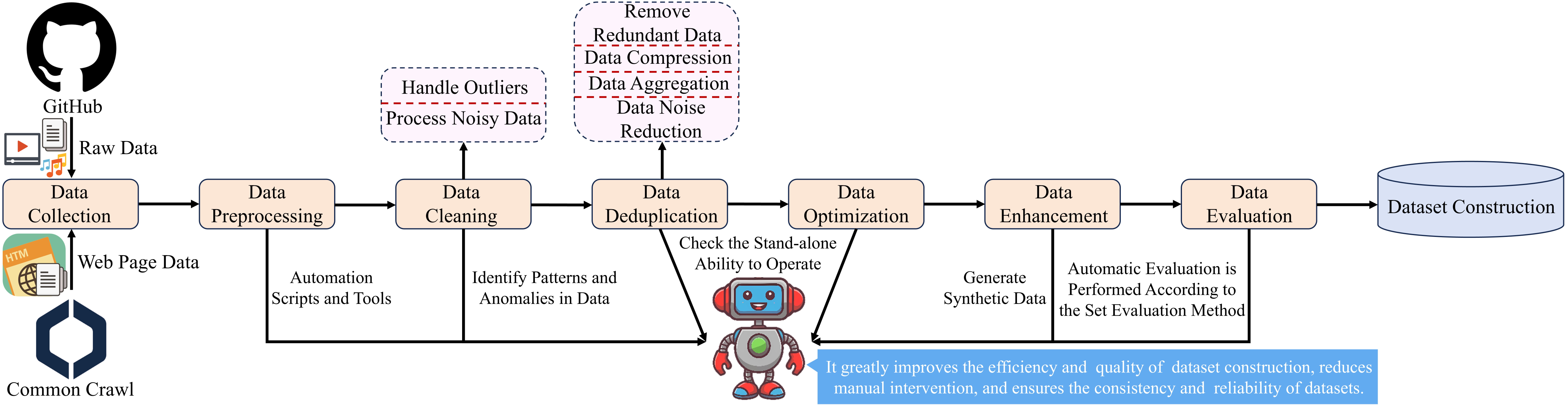}
    \caption{The dataset building process.}
    \label{fig:fig55}
\end{figure*}

In recent years, the application of reinforcement learning algorithms to datasets has matured \cite{levine2020offline}. However, the trial-and-error learning and reward mechanisms of reinforcement learning rely on interactions between the agent and the environment, which are often time-consuming and costly. The emergence of offline reinforcement learning algorithms offers the possibility of data-driven approaches without requiring expensive real-world exploration, instead relying on large pre-collected datasets. Offline reinforcement learning methods can provide effective initialization for online fine-tuning, but current benchmark tasks for evaluating offline reinforcement learning algorithms are relatively simple and approaching performance saturation. The advent of the D5RL dataset provides a new perspective for evaluating offline reinforcement learning. It offers both offline and online fine-tuning evaluations and designs specific pretraining and fine-tuning for certain tasks. By simulating accessible evaluation environments and providing a level of realism that reflects real-world system attributes, D5RL lays a new foundation for the development of offline reinforcement learning datasets \cite{rafailov2024d5rl}. Figure \ref{fig:fig55} demonstrates the process of constructing a dataset, ranging from data collection and preprocessing to data optimization and augmentation, with automation and efficiency fully achieved in the process. 

\subsubsection{Pre-training}
During the pre-training for Code LLMs, datasets often comprise a large-scale code corpus along with a smaller fraction of data sources like mathematics, text, etc., or involve fine-tuning a code corpus on top of general-purpose LLMs. Code LLM architectures, similar to generic LLMs, can be categorized into three types: encoder-only, decoder-only, and encoder-decoder models. Encoder-only models, such as CodeBERT, are generally effective for code understanding tasks, including type prediction, code retrieval, and code clones detection \cite{feng2020codebert}. Decoder-only models, like StarCoder, excel in generative tasks such as code generation, translation, and summarization \cite{li2023starcoder}. Encoder-decoder models, such as CodeT5, are capable of addressing both code understanding and generation tasks, although they are not necessarily superior to their encoder-only or decoder-only counterparts \cite{wang2021codet5}. Additionally, some Code LLMs, like Qwen2.5-Coder, leverage synthetic data in their training procedures, thereby exhibiting higher comprehension abilities \cite{hui2024qwen2}.

Within the field of code generation using LLMs, the design of current models generally belongs to one of the two main types: encoder-decoder architectures, like CodeT5, CodeT5+, and CodeRL \cite{ye2019understanding,wang2021codet5,wang2023codet5+,le2022coderl}; or decoder-only architectures, such as Codex, StarCoder, CodeLlama, and CodeGemma \cite{wu2023decoder,li2023starcoder,roziere2023code,team2024codegemma}.

\subsection{Post-training}
The searching phase in post-training is essential for refining Code LLMs by systematically exploring and optimizing solutions within large and complex code spaces. This stage involves generating candidate solutions using pre-trained models through probabilistic decoding methods such as beam search, Top-k sampling, nucleus sampling, or deterministic strategies like greedy search, ensuring a balance between diversity and high-confidence outputs. To evaluate these candidates, functional correctness is prioritized using metrics like pass@k and unit tests, which directly assess program functionality and complement traditional token similarity measures. Reinforcement learning techniques, such as Proximal Policy Optimization (PPO), are often employed to iteratively refine candidates by leveraging execution feedback and semantic evaluations, thereby optimizing policies to align with syntactic and functional criteria. Given the computational intensity of searching, strategies like model quantization, caching, and parallelized execution pipelines are employed to enhance scalability and efficiency. Overall, this phase bridges the gap between pre-trained model capabilities and real-world requirements, ensuring that the generated code is both high-quality and functional for practical applications.

OpenCoder significantly improves its proficiency in theoretical computer science and practical programming tasks through a two-phase instruction fine-tuning process, overcoming the limitations of models concentrating on just one field. DeepSeekMath uses Online Rejection Sampling Fine-tuning (Online RFT) which differs from conventional methods by leveraging outputs from a live policy model for fine-tuning, rather than a supervised approach. Similarly, Group Relative Policy Optimization (GRPO), a variation of Proximal Policy Optimization (PPO), enhances policies by evaluating relative rewards from multiple outputs for the same problem, instead of depending on a single value function. These methods strive to enhance the model's problem-solving abilities through more dynamic and context-sensitive learning processes. Using a framework known as CORGI (Controlled Generation with RL for Guided Interaction), certain researchers have enabled models to obtain immediate textual feedback over several cycles. This is accomplished by simulating interactive sessions with an automated critique system, prompting the models to modify their responses to adhere to predefined constraints derived from the feedback. Qwen2.5-Coder series utilizes a comprehensive strategy to improve model performance. It involves creating instruction-tuning datasets by identifying multilingual programming code and generating instructions from GitHub. A combined method of coarse-to-fine tuning and mixed tuning strategy then incorporates both low- and high-quality instruction samples, enhancing the model's ability to respond to commands. Furthermore, data decontamination is performed to reduce test set leakage effects on evaluation accuracy. Together, these approaches enhance the robustness and effectiveness of the Qwen2.5-Coder series for coding tasks.

In the post-training phase, preference feedback-based learning methods, especially PPO (Proximal Policy Optimization) and DPO (Direct Preference Optimization), have become the key techniques to improve the performance of language models. PPO, as an online reinforcement learning algorithm in the post-training phase, includes reward model training and policy model optimization. First, PPO uses the preference data to train a reward model, which evaluates the quality of the responses generated by the model and becomes the objective function for subsequent policy optimization. Next, PPO leverages the reward model to score the responses generated by the strategy model and further uses this score to optimize the strategy model. This optimization process ensures training stability by introducing a KL penalty term that prevents the model's policy distribution from deviating too much from the initial policy. The advantage of PPO is that its ability to train with online data helps the model to remain exploratory and adaptable in real-world applications, especially for tasks that require complex reasoning and coding capabilities. However, the online training approach of PPO brings higher computational cost and engineering complexity. In contrast, DPO, as an offline reinforcement learning approach, eliminates the step of training reward models by optimizing policy models directly on preference data, which simplifies the training process and effectively reduces the demand for computational and engineering resources. The optimization process of DPO improves the performance of the policy model by increasing the log-likelihood difference between the selected and rejected responses while ensuring that the model does not overly deviate from the initial strategy. DPO is computationally efficient and is particularly suitable for resource-constrained environments, although it may not be as good as PPO in terms of model adaptability and exploration capabilities.

Comparing PPO and DPO, each has its own strengths and limitations in the post-training phase. PPO shows stronger potential and typically performs better on multiple tasks, especially on tasks involving complex reasoning and coding capabilities. However, DPO is ideal for resource-constrained environments due to its more streamlined training process and lower computational cost. Overall, PPO and DPO provide two different paths for language model optimization in the post-training phase, with PPO providing stronger performance on complex tasks, while DPO meets more stringent resource constraints by improving computational efficiency. As language modeling technology continues to evolve, the selection of an appropriate post-training method will depend on specific application requirements, resource conditions, and performance goals.

\section{Related Applications of Reinforcement Learning}
\subsection{Enhancing Code Language Models with Reinforcement Learning}
The integration of reinforcement learning (RL) into code language models (Code LLMs) offers a significant enhancement in accurate and efficient code generation through interactive feedback mechanisms. RL frameworks effectively handle vital aspects such as unit tests and functional correctness by employing real-time evaluation and iterative improvement methods that traditional supervised fine-tuning often overlooks. CodeRL \cite{13} is an example that combines pre-trained language models with deep RL to refine code generation processes. In this system, the code model acts as an actor network, while a critic network evaluates the functional correctness of the generated code, using feedback from unit tests as reward signals. This interactive training cycle continuously enhances the syntactic and semantic precision of the model, producing code that is both accurate and robust particularly in complex programming settings. Further advancements in RL-based code generation are demonstrated by PPOCoder \cite{8} and PanGu-Coder2 \cite{6}. PPOCoder merges CodeT5 with Proximal Policy Optimization (PPO), improving stability and reliability by restricting policy updates and using execution feedback to optimize the code's structure and function. A reward function assesses the alignment between the code's Abstract Syntax Tree (AST) and the ground truth, enabling PPOCoder to generate highly precise, functionally correct code. On the other hand, PanGu-Coder2 uses Ranking Reinforcement from Human Feedback (RRHF) to directly integrate human preferences into the code generation process. This framework employs ranking-based reinforcement to emphasize outputs that satisfy human expectations, considerably boosting the relevance and quality of code generated for complex tasks. Collectively, these frameworks demonstrate how RL-enhanced Code LLMs can dynamically evolve to achieve excellence in code functionality, accuracy, and alignment with human standards.

\subsection{The Impact of Reinforcement Learning on End-to-end Software Development}

Reinforcement learning (RL) is revolutionizing software engineering, contributing to enhancements across various phases from design to deployment \cite{Zhuang2021A}. By learning optimal actions through environmental interactions, RL facilitates automation in areas such as code suggestion and generation, accelerating development and minimizing human error. In the realm of software testing, RL streamlines test case generation, determines execution orders, and optimizes processes, thus enhancing test quality and coverage. For network control, RL also advances traffic management and control strategies, which are essential for distributed systems \cite{Xiao2021Leveraging}. 

As RL technology advances, its role in comprehensive software development—spanning code creation, testing, and resource management—will become increasingly prominent, establishing it as an indispensable tool for software engineers. GitHub Copilot, utilizing OpenAI’s Codex model, employs Reinforcement Learning from Human Feedback (RLHF) to improve functions such as code completion, generation, refactoring, and documentation. This strategy, in conjunction with extensive training on large code datasets, enables Copilot to deliver real-time coding support in popular IDEs like Visual Studio and JetBrains with substantially boosted developing efficiency. Similarly, Zhipu AI’s CodeGeeX, leveraging the ChatGLM model with RLHF, supports code generation, translation, and completion across various languages in IDEs including VS Code and IntelliJ. Huawei’s CodeArts Snap, using PanGu-Coder2 and Reinforcement Learning from Human Reverse Feedback (RRHF), enhances code generation, debugging, and test generation with contextually tailored code recommendations. These models, optimized through RLHF or RRHF, exhibit effective end-to-end applications by aligning code generation with actual developer needs in IDE settings.

\section{Metrics and Benchmarks}
\subsection{Metrics}
Identifying efficient and reliable automatic evaluation metrics for code generation has been a significant challenge \cite{Ren2020CodeBLEUAM}. Initially, drawing inspiration from machine translation and text summarization, many efforts relied on metrics that evaluated token matching. Notable examples include BLEU \cite{papineni-etal-2002-bleu}, ROUGE \cite{lin-2004-rouge}, and METEOR \cite{banerjee-lavie-2005-meteor}. Nevertheless, these methods typically struggle to accurately assess the syntactic and functional accuracy of the code, as well as its semantic attributes. Furthermore, these metrics are not tailored for various programming languages and specific compilers, which also diminishes their practicality. To mitigate these limitations in token-matching based metrics, CodeBLEU \cite{Ren2020CodeBLEUAM} was developed. It combines syntactic and semantic elements from Abstract Syntax Trees (ASTs) and Data Flow Graphs (DFGs) with conventional BLEU scores, thereby enhancing the evaluation precision for code generation. However, CodeBLEU still fails to fully resolve the issues related to execution errors and discrepancies in execution results. 

Given these obstacles, execution-based metrics have become increasingly important for assessing code generation. Notable methodologies include execution accuracy \cite{Rajkumar2022EvaluatingTT}, pass@t \cite{olausson2024selfrepairsilverbulletcode}, n@k \cite{doi:10.1126/science.abq1158}, and pass@k \cite{Chen2021EvaluatingLL}. When reinforcement learning (RL) is applied to enhance the code generated by large language models (LLM), execution-based metrics, especially pass@k \cite{Chen2021EvaluatingLL}, have shown greater significance. The estimation of pass@k is described as follows:
\begin{equation}
    \text{pass@k} := \mathbb{E}_{\text{problems}} \left[ 1 - \frac{\binom{n-c}{k}}{\binom{n}{k}} \right]
\end{equation}

 \noindent Here, $c$ represents the number of successful tests among the generated $n$ codes, with a larger $n$ resulting in a more precise estimate. It assesses the likelihood that at least one of the created code samples $k$ passes all unit tests. Such metrics are crucial in establishing the functional correctness of the generated code by examining its execution performance, serving as a key evaluation tool for modern Code LLMs. For an evaluation of the code generated by Code LLMs enhanced with reinforcement learning, see Table \ref{table:1}. However, these execution-focused evaluation techniques depend heavily on the integrity of unit tests and are confined to estimating the quality of executable code. In scenarios where unit tests are not suitable, token matching metrics are frequently employed as an alternative evaluation approach. 

In summary, choosing the right metrics to assess the quality of code generated by Code LLMs is essential. While current methods, such as token-matching and execution-based approaches, are well established, there remains a shortage of metrics to effectively evaluate the generated code's security and efficiency. More sophisticated metrics are necessary to evaluate the models.

\begin{table*}[ht]
\centering
\caption{ Performance of Base, Reinforcement Learning-based, and Multi Reinforcement Learning-based Models on \textbf{APPS} benchmark}

\resizebox{\linewidth}{!}{  
\begin{tabular}{p{2.8cm}p{1.8cm}p{1cm}p{1cm}p{1cm}p{1cm}p{1cm}p{1cm}p{1cm}p{1cm}p{1cm}p{1cm}p{1cm}p{1cm}}
\toprule
\hline
\multirow{2}{*}{\textbf{Model}} & \multirow{2}{*}{\textbf{Size}} & \multicolumn{4}{c}{\textbf{Pass@1}} & \multicolumn{4}{c}{\textbf{Pass@5}} & \multicolumn{4}{c}{\textbf{Pass@1000}} \\
\cmidrule(lr){3-6} \cmidrule(lr){7-10} \cmidrule(lr){11-14}
& & Intro & Inter & Comp & All & Intro & Inter & Comp & All & Intro & Inter & Comp & All \\
\midrule
\multicolumn{14}{c}{\textbf{Base Models}} \\  
\midrule 
Codex & 12B & 4.14 & 0.14 & 0.02 & 0.92 & 9.65 & 0.51 & 0.09 & 2.25 & 25.02 & 3.70 & 3.23 & 7.87 \\
AlphaCode & 1B & - & - & - & - & - & - & - & - & 17.67 & 5.24 & 7.06 & 8.09 \\
GPT-3 & 175B & 0.20 & 0.03 & 0.00 & 0.06 & 2.70 & 0.73 & 0.00 & 1.02 & - & - & - & - \\
GPT-2 & 0.1B & 1.00 & 0.33 & 0.00 & 0.40 & 3.60 & 1.03 & 0.00 & 1.34 & - & - & - & - \\
GPT-2 & 1.5B & 1.30 & 0.70 & 0.00 & 0.68 & 5.50 & 1.03 & 0.00 & 1.58 & 27.90 & 9.27 & 8.80 & 12.32 \\
GPT-Neo & 2.7B & 3.90 & 0.57 & 0.00 & 1.12 & 5.50 & 0.80 & 0.00 & 1.58 & 27.90 & 9.83 & 11.40 & 13.76 \\
GPT-J & 6B & 5.60 & 1.00 & 0.50 & 1.82 & 9.20 & 1.73 & 1.00 & 3.08 & 35.20 & 13.15 & 13.51 & 17.63 \\
CodeT5†  & 60M & 1.40 & 0.67 & 0.00 & 0.68 & 2.60 & 0.87 & 0.10 & 1.06 & - & - & - & -\\
CodeT5†  & 220M & 2.50 & 0.73 & 0.00 & 0.94 & 3.30 & 1.10 & 0.10 & 1.34 & - & - & - & -\\
CodeT5†  & 770M & 3.60 & 0.90 & 0.20 & 1.30 & 4.30 & 1.37 & 0.20 & 1.72 & - & - & - & - \\
\midrule
\multicolumn{14}{c}{\textbf{Reinforcement Learning-based Models}} \\  
\midrule 
CodeRL & 770M & 6.20 & 1.50 & 0.30 & 2.20 & 9.39 & 1.90 & 0.42 & 3.10 & 35.30 & 13.33 & 13.60 & 17.78 \\
PPOCoder    & 770M & 5.20 & 1.00 & 0.50 & 1.74 & 9.10 & 2.50 & 1.20 & 3.56 & 35.20 & 13.35 & 13.90 & 17.77 \\
RLTF & 770M & 4.16 & 0.97 & 0.20 & 1.45 & 10.12 & 2.65 & 0.82 & 3.78 & 38.30 & 15.13 & 15.90 & 19.92 \\
$\mathcal{B}$-Coder & $\leq$770M/stage$^3$ & 6.70 & 1.50 & 0.30 & 2.30 & 10.40 & 2.63 & 0.70 & 3.80 & 37.00 & 13.67 & 12.60 & 18.12 \\

\midrule
\multicolumn{14}{c}{\textbf{Multi Reinforcement Learning- based Models}} \\ 
\midrule 
CodeRL + CodeT5 & 770M & 4.90 & 1.06 & 0.5 & 1.71 &8.60 & 2.64 & 1.0 & 3.51 & 36.10 & 12.65 & 13.48 & 17.50 \\
PPOCoder + CodeT5 & 770M & 5.20 & 1.00 & 0.5 & 1.74 &9.10 & 2.50 & 1.20 & 3.56 & 35.20 &13.35 & 13.90 & 17.77 \\
\hline
\bottomrule
\end{tabular}
}
\label{table:1}
\end{table*}

\subsection{Benchmarks}
To thoroughly evaluate the performance of large language models (LLMs) in code generation, researchers have recently developed numerous high-caliber benchmarks. Building upon foundational studies, several variations of the HumanEval dataset have been introduced, alongside additional benchmarks designed to assess the code generation abilities of LLMs in a wider scope. Benchmarks used for testing, often involving reinforcement learning-related Code LLMs, typically encompass the following elements.

HumanEval features 164 selected Python programming tasks, each including a function signature, a descriptive docstring, an implementation, and several unit tests \cite{zheng2023codegeex}. HumanEval+ expands on the original HumanEval benchmark by increasing the number of test cases by 80-fold. This expanded testing capability allows HumanEval+ to detect a significant amount of previously unnoticed flawed code produced by LLMs \cite{liu2024your}. MBPP is a collection of around 974 beginner-level Python coding tasks sourced from public contributions. Each task offers an English description, a code solution, and three automated test cases. MBPP+ enhances MBPP by removing poorly designed problems and fixing flawed solutions. It also boosts the test capacity by a factor of 35 to improve coverage \cite{guo2023instruction}. 

In the realm of competitions, the APPS benchmark contains 10,000 Python problems spanning three levels of difficulty: introductory, interview, and competition. Each problem includes a description in English, a correct Python solution, and corresponding test cases defined by inputs and outputs or function names, if available. The APPS+ dataset consists of 7,456 entries. It improves upon the original APPS dataset by eliminating defective entries, standardizing input and output formats, and ensuring quality and coherence through unit tests and manual review. Each entry includes a problem description, a standard solution, a function name, unit tests, and initial code. LiveCodeBench provides a comprehensive and uncontaminated benchmark designed to assess a wide range of coding skills in LLMs, such as code creation, self-repair, execution, and test output prediction \cite{jain2024livecodebench}. It continuously gathers new coding challenges from competitions on three renowned platforms: LeetCode, AtCoder, and CodeForces. The dataset's latest update includes 713 problems released between May 2023 and September 2024.

\section{Prospects for Future Development}
Through our survey, it is evident that reinforcement learning (RL) emerges as a transform strategy for enhancing large language code models (Code LLMs) in the domain of code generation with significant advancements in performance. However, there are still several challenges that require addressing:

\paragraph{Creating High-Quality Code Datasets} 
The success of Code LLMs is greatly influenced by the diversity and quality of the code datasets used for pretraining and fine-tuning. At present, there is a shortage of comprehensive, high-quality datasets that cover a wide array of programming tasks, styles, and languages. This shortfall impedes the ability of LLMs to generalize to new programming tasks, various coding settings, and real-world software development. Advanced data acquisition methods, including automated mining of code repositories, sophisticated filtering techniques, and code data synthesis, could facilitate the development of more enriched datasets. Although RL algorithms often excel at specific tasks, their challenge lies in adapting to new tasks or environments, which restricts their versatility and applicability.

\paragraph{Formulating Comprehensive Benchmarks and Metrics for Code Generation in Code LLMs}
Current benchmarks, such as HumanEval, may not comprehensively evaluate the array of coding skills required in practical software development. Additionally, many of the evaluation metrics currently in use prioritize syntactic accuracy or functional performance, overlooking critical aspects such as code efficiency and security. Creating benchmarks that replicate the complexities of real-world software development could provide a more accurate evaluation of the coding ability of LLMs.

\paragraph{Enhancing Support for Low-Level and Domain-Specific Programming Languages}
LLMs are mainly trained on popular high-level languages, resulting in limited support for low-level and domain-specific languages like assembly and lean. This underrepresentation curtails the use of LLMs in specialized domains and systems programming. Progressing research in transfer learning and meta-learning could allow LLMs to apply knowledge from widely-used languages to improve their performance on lesser-known ones.

\paragraph{Minimizing Computational Costs}
RL algorithms, especially those involving extensive state spaces or intricate decision-making processes, typically require significant computational power, such as high-performance GPUs and substantial memory. Such demands can be prohibitive in environments with limited resources. Exploring more efficient RL methods and refining resource utilization can help mitigate these computational needs. Confronting these challenges is essential for fully realizing the potential of RL-augmented LLMs in code generation and enhancing their capabilities across varied programming domains.

\section{Conclusion}
In this paper, we discuss current reinforcement learning (RL) approaches to code generation and optimization, and analyze various RL-based strategies in different generation and optimization directions. We examine the commonalities and differences of these methods, arguing that RL holds significant promise for code generation and optimization, potentially marking a major shift in the field. Our goal is to help researchers gain a comprehensive understanding of the possible directions and the core challenges, and to inspire future advancements and progresses in this evolving field.
      
\section{Limitations}
In this comprehensive survey, we have examined the application of reinforcement learning in code generation and optimization, analyzing a range of methods and techniques. However, due to space limitations, we have not provided a comprehensive analysis of all aspects under discussion. Firstly, we did not provide a detailed account of the datasets employed for model training, which are of paramount importance for the model's generalization and performance. Further research could examine the influence of disparate datasets on model performance and the construction of more diverse and representative datasets to improve generalization. Secondly, the scalability and generalization of reinforcement learning models in the context of large-scale codebases and across multiple projects were not discussed. In practical applications, models must be capable of handling codebases of varying scales and complexities, necessitating good scalability and adaptability. Further research could concentrate on improving the scalability of the models and the transfer of learning between disparate projects and programming languages. Lastly, a detailed comparison of the training and inference times of various algorithms was not provided. In the context of software development, the efficiency of the algorithm is of paramount importance. Consequently, future studies could assess the time complexity of different reinforcement learning algorithms during the training and inference phases to optimize these times.

\bibliography{custom}

\appendix

\end{document}